\documentclass[preprint,preprintnumbers,amssymb,aps]{revtex4}
\usepackage{hyperref}
\usepackage{floatrow}
\usepackage{graphicx}
\usepackage{dcolumn}
\usepackage{amsmath}
\usepackage{epsfig}
\usepackage{subfig}
\usepackage{array}
\usepackage{epstopdf} 
\usepackage{xcolor}
\usepackage{bm}

\def\be{\begin{equation}}
\def\ee{\end{equation}}
\def\bea{\begin{eqnarray}}
\def\eea{\end{eqnarray}}
\def\NO{\nonumber}
\def\gev{\mathrm{~GeV}}

\begin{document}

\normalsize

\title{$\eta_c$ production associated with light hadrons at the B-factories and the future super B-factories}

\author{Qin-Rong~Gong$^{1}$}
\author{Zhan~Sun$^{2}$}
\author{Hong-Fei~Zhang$^{3}$}
\author{Xue-Mei Mo$^{4}$}
\affiliation{
\footnotesize
\it $^{1}$Department of Physics and State Key Laboratory of Nuclear Physics and Technology, Peking University, Beijing 100871, P. R. China \\
\it $^{2}$School of Science, Guizhou Minzu University, Guiyang 550025, P. R. China \\
\it $^{3}$School of Science, Chongqing University of Posts and Telecommunications, Chongqing 400065, P. R. China \\
\it $^{4}$Institute of Digital Medcine, School of Biomedical Engineering, Third Military Medical University, Chongqing, 400038, China
}

\begin{abstract}

We present a complete study of the associated production of the $\eta_c$ meson with light hadrons in $e^+e^-$ collisions at the B-factory energy,
which is demonstrated to be one of the best laboratories for testing the colour-octet (CO) mechanism.
The colour-siglet contributions are evaluated up to $O(\alpha^2\alpha_s^3)$ while the CO ones are evaluated up to $O(\alpha^2\alpha_s^2)$.
For the first time, the angular distribution of the $^1S_0^{[8]}$ production is studied at QCD next-to-leading order.
We find that the $^1S_0^{[8]}$ channel dominates the total cross section,
while the $^1P_1^{[8]}$ one exhibits its importance in the angular distribution,
which turns out to be downward going with respect to $\mathrm{cos}\theta$.
This can be considered as the most distinct signal for the CO mechanism.

\end{abstract}

\maketitle

\section{Introduction}\label{intro}

$\eta_c$ ($0^{-+}$), known as the lightest charmonium state,
can provide a very good laboratory for the study of the quarkonium production mechanism.
However, in contrast to the copious data on the $J/\psi$ yield,
the observation of the $\eta_c$ meson is scant.
This is basically because the $J/\psi$ can be detected via its leptonic decay channels,
while the fragments of the $\eta_c$ decays are dominated by multiple hadrons~\cite{Agashe:2014kda},
both the observation and reconstruction of which are more difficult.
A novel approach to the measurement of the various charmonium states using their common decay channel to $p\bar{p}$ was proposed in Ref.~\cite{Barsuk:2012ic},
which shed light on the investigation of the $\eta_c$ and $h_c$ mesons.
By exploiting this approach, LHCb Collaboration~\cite{Aaij:2014bga} achieved their first study on the inclusive and prompt $\eta_c$ yield in $pp$ collisions.
They found that the $\eta_c$ hadroproduction cross section is even larger than that of the $J/\psi$ in the same experimental condition.
On the theory side, QCD leading order (LO) calculation of the $\eta_c$ hadroproduction within the nonrelativistic QCD (NRQCD)~\cite{Bodwin:1994jh}
was accomplished in Refs.~\cite{Biswal:2010xk, Likhoded:2014fta},
following which the complete QCD next-to-leading order (NLO) studies came out in a few weeks~\cite{Butenschoen:2014dra, Han:2014jya, Zhang:2014ybe}.
Ref.~\cite{Butenschoen:2014dra} considered the LHCb data on $\eta_c$ hadroproduction as the challenge to NRQCD,
while Refs.~\cite{Han:2014jya, Zhang:2014ybe} found these data did not bring in any inconsistency.
Ref.~\cite{Zhang:2014ybe} further argued that this measurement actually provided an excellent opportunity for fixing the $\eta_c$ (as well as the $J/\psi$) wave function at the origin,
and can also help with the determination of the colour-octet (CO) long-distance matrix elements (LDMEs) for the $J/\psi$ production.
As was pointed out in Ref.~\cite{Ma:2010yw}, only two degrees of freedom of the three $J/\psi$ CO LDMEs can be fixed by the $J/\psi$ yield data.
$\eta_c$ data helped to fix the last one, $\langle O^{J/\psi}(^1S_0^{[8]})\rangle$.
Having this parameter fixed, Ref.~\cite{Sun:2015pia} discovered some interesting features of the $J/\psi$ hadroproduction and polarization,
which provided a possibility for the solution to the long-standing $J/\psi$ polarization puzzle.

In fact, as early as 17 years ago, $\eta_c$ photo- and leptoproduction as a heuristic probe to the CO mechanism has already been proposed~\cite{Hao:1999kq, Hao:2000ci}.
In these processes, the colour-siglet (CS) channel is suppressed by an order of $\alpha_s^2$ compared to the $^1S_0^{[8]}$ channel,
which, on the one hand, provided an opportunity to test the CO mechanism,
on the other hand, could help to fix the $^1S_0^{[8]}$ LDME for $\eta_c$ production.
Unfortunately, due to lack of data, this device has never been put into implementation.

Similar to the $\eta_c$ photo- and leptoproduction, $\eta_c$ production in $e^+e^-$ annihilation also has these good features.
This process becomes more important since the B-factories raised up their luminosity to the order of $10^{34}\mathrm{cm^{-2}s^{-1}}$ ($10^{-2}\mathrm{pb}^{-1}\mathrm{s}^{-1}$).
Two super B-factories~\cite{Biagini:2011zz, Abe:2010gxa} are proposed to reach even higher luminosities,
on the order of $10^{36}\mathrm{cm^{-2}s^{-1}}$  ($1\mathrm{pb}^{-1}\mathrm{s}^{-1}$).
A few years running of these machines can accumulate adequate data for a precision measurement of the $\eta_c$ production,
which, as will be shown later, can provide the most distinct test of NRQCD.

For the $\eta_c$ production, up to $v^4$,
one CS state ($^1S_0^{[1]}$) and three CO states ($^1S_0^{[8]}$, $^3S_1^{[8]}$ and $^1P_1^{[8]}$) are involved.
At the B-factory energy, charge parity is approximately conserved.
The CS state can only be produced with at least three gluons emitted.
This process is of order $\alpha^2\alpha_s^3$.
However, $^1S_0^{[8]}$ state can be produced with only one gluon emitted,
which is of order $\alpha^2\alpha_s$, two orders lower than the CS one in $\alpha_s$.
We can expect the CO processes be more significant than the CS one.
Thus the measurement can definitely distinguish the two mechanisms.

Another interesting feature of this process is that,
in constrast to the $\eta_c$ hadroproduction case in which the $^3S_1^{[8]}$ channel dominate the production,
$\eta_c$ production in $e^+e^-$ annihilation is dominated by the $^1S_0^{[8]}$ and $^1P_1^{[8]}$ channels.
The LDMEs for these channels are related to the $^3S_1^{[8]}$ and $^3P_J^{[8]}$ LDMEs for the $J/\psi$ production by the heavy quark spin symmetry (HQSS).
Since the determination of the $J/\psi$ LDMEs is still facing controversy
~\cite{Butenschoen:2011yh, Chao:2012iv, Gong:2012ug, Bodwin:2014gia, Faccioli:2014cqa, Butenschoen:2014dra, Han:2014jya, Zhang:2014ybe},
this process can help to clarify this issue.

The last but not the least important thing to mention:
the measurement of the $\eta_c$ production at B-factories might provide some useful information for the study of the process $e^+e^-\rightarrow J/\psi+X$,
which was measured by BABAR~\cite{Aubert:2001pd} and Belle~\cite{Abe:2001za, Abe:2002rb, Pakhlov:2009nj} Collaborations.
The theoretical studies of these experiments are presented in Refs.~\cite{Zhang:2006ay, Ma:2008gq, Gong:2009kp, Gong:2009ng, He:2009uf, Shao:2014rwa},
which found that the CS results of the total cross sections generally saturate the most recent Belle measurement~\cite{Pakhlov:2009nj},
and the inclusion of the CO contributions would ruin the agreement between the theory and experiment.
In spite of this, the angular distribution~\cite{Gong:2009ng}, within the CS mechanism,
for the production of the $J/\psi$ in association with either light hadrons or charmed hadrons is in conflict with the data given in the same experiment paper.
Ref.~\cite{Zhang:2014ybe} suggested that the CS LDME for the $J/\psi$ production might be smaller than the ordinarily used values
obtained in potential-model calculations~\cite{Eichten:1995ch},
which left room for the CO mechanism.
The smaller CS LDME and the inclusion of the CO contributions can provide opportunities for the understanding of the angular distribution puzzle.
However, the results given by employing the LDMEs in Ref.~\cite{Zhang:2014ybe} exceed the Belle measurement of the production of the $J/\psi$ plus light hadrons.
This problem is still waiting for further investigation.
Actually, many factors can cause this discrepancy.
For example, the $\alpha_s^2$ corrections are always significant~\cite{Beneke:1997jm, Chen:2015csa, Feng:2015uha},
thus the universality of the LDMEs at QCD NLO can not take the responsibility of testing NRQCD.
Before we can achieve the high-order calculations, $\eta_c$ production at B factories can serve as an alternative test of the CO mechanism.
Since this process is dominated by the $^1S_0^{[8]}$ and $^1P_1^{[8]}$ channels,
the measurement can, on the one hand, distinguish the CS and CO contributions, on the other hand,
specify whether the theoretical results for the $c\bar{c}(^1S_0^{[8]})$ production in $e^+e^-$ annihilation reach a good convergence up to QCD NLO.

In this paper, we study the $\eta_c$ associated production with light hadrons at B-factory energy within the NRQCD framework,
which can provide references for the future experiment at the Super B-factories.
The rest of this paper is organised as follows.
In section II, we briefly describe the framework of our calculation.
Section III presents the numerical results and discussions,
while we come to our conclusions in section IV.

\section{$\eta_c$ Associated Production with Light Hadrons within the NRQCD Framework }

In the NRQCD factorization framework, up to $v^4$, four intermediate $c\bar{c}$ states,
including one CS state ($^1S_0^{[1]}$) and three CO states ($^1S_0^{[8]}$, $^3S_1^{[8]}$ and $^1P_1^{[8]}$), are involved in the $\eta_c$ production.
The cross section for the $\eta_c$ production in association with light hadrons in $e^+e^-$ collisions can thus be expressed as
\be
d\sigma(e^+e^-\rightarrow\eta_c+X)=\sum_nd\hat{\sigma}(e^+e^-\rightarrow c\bar{c}(n)+X)\langle O^{\eta_c}(n)\rangle, \label{eqn:nrqcdfac}
\ee
where $n$ runs over the four intermediate states,
$\hat{\sigma}$ are the corresponding short-distance coefficients (SDCs),
and $X$ denotes light hadrons, the hadronization process of which are not concerned in our calculation.
Thus, we simply evaluate the processes in which $X$ are partons (gluons and/or light quarks).

The charge parity of the CS state, $^1S_0^{[1]}$, is $+1$.
Since the charge parity is conserved in strong and electroweak interactions,
this state can be produced with at least three gluons emitted.
However, the LO processes for $^1S_0^{[8]}$, $^3S_1^{[8]}$ and $^1P_1^{[8]}$ productions involve only one, two and two emitted gluons, respectively.
This results in the fact that the CS contribution are greatly suppressed compared with the CO one.
In this paper, we consider the CS contribution at LO ($\alpha^2\alpha_s^3$).
There is only one process at this order, namely
\be
e^+e^-\rightarrow c\bar{c}(^1S_0^{[1]})+g+g+g, \label{eqn:cs}
\ee
where $g$ denotes a gluon.

The CO processes are evaluated up to the order $\alpha^2\alpha_s^2$,
which is, for both the $^3S_1^{[8]}$ and $^1P_1^{[8]}$ channels, LO,
while for the $^1S_0^{[8]}$ channel, NLO in $\alpha_s$.
The processes involved are
\bea
&&e^+e^-\rightarrow c\bar{c}(^1S_0^{[8]})+g, \label{eqn:1s08lo} \\
&&V: e^+e^-\rightarrow c\bar{c}(^1S_0^{[8]})+g, \label{eqn:1s08v} \\
&&e^+e^-\rightarrow c\bar{c}(^1S_0^{[8]})+g+g, \label{eqn:1s08gr} \\
&&e^+e^-\rightarrow c\bar{c}(^1S_0^{[8]})+q+\bar{q}, \label{eqn:1s08qr} \\
&&e^+e^-\rightarrow c\bar{c}(^3S_1^{[8]})+g+g, \label{eqn:3s18g} \\
&&e^+e^-\rightarrow c\bar{c}(^3S_1^{[8]})+q+\bar{q} \label{eqn:3s18q} \\
&&e^+e^-\rightarrow c\bar{c}(^1P_1^{[8]})+g+g, \label{eqn:1p18}
\eea
where $q$ and $\bar{q}$ represent light quark and antiquark, respectively,
and the label $V$ means one-loop-level virtual correction to the process on the right-hand side of it.
Summing over the processes in Eq.(\ref{eqn:1s08v}), Eq.(\ref{eqn:1s08gr}) and Eq.(\ref{eqn:1s08qr}),
the cross section for the $\eta_c$ production via the $^1S_0^{[8]}$ channel at QCD NLO will be free of divergence,
while those for the processes listed in Eq.(\ref{eqn:cs}), Eq.(\ref{eqn:1s08lo}), Eq.(\ref{eqn:3s18g}) and Eq.(\ref{eqn:3s18q}) are nonsigular in theirselves.
However, the process in Eq.(\ref{eqn:1p18}) is divergent.
This divergence can be cancelled within the NRQCD framework by including the QCD corrections to the $^1S_0^{[8]}$ LDME.
Summing the two contributions stated above, we can redefine the $^1P_1^{[8]}$ SDC as a finite quantity.
The detail of this procedure can be found in Ref.~\cite{Wang:2012tz, Wang:2014vsa, Jia:2014jfa},
so, we just omit these discussions in the current paper,
and purloin the useful equations in the references.
One important feature necessary for our discussion is that the SDC for the $^1P_1^{[8]}$ channel can be decomposed in two parts
\be
\hat{\sigma}(e^+e^-\rightarrow c\bar{c}(^1P_1^{[8]})+g+g)=\hat{\sigma}_{foml}
-\frac{\alpha_s}{9\pi m_c^2}\frac{N_c^2-4}{N_c}\mathrm{ln}(\frac{\mu_\Lambda^2}{m_c^2})\hat{\sigma}(e^+e^-\rightarrow c\bar{c}(^1S_0^{[8]})+g),
\ee
where $\hat{\sigma}_{foml}$ is completely free of $\mu_\Lambda$, the NRQCD factorisation scale.

Then we rewrite Eq.(\ref{eqn:nrqcdfac}), up to the order we maintain in our calculation, in an explicit form as
\bea
d\sigma(\eta_c)&=&d\hat{\sigma}(^1S_0^{[1]})\langle O^{\eta_c}(^1S_0^{[1]})\rangle
+d\hat{\sigma}(^3S_1^{[8]})\langle O^{\eta_c}(^3S_1^{[8]})\rangle+d\hat{\sigma}_{foml}\langle O^{\eta_c}(^1P_1^{[8]})\rangle \NO \\
&-&\frac{\alpha_s}{9\pi m_c^2}\frac{N_c^2-4}{N_c}\mathrm{ln}(\frac{\mu_\Lambda^2}{m_c^2})d\hat{\sigma}_{lo}(^1S_0^{[8]})\langle O^{\eta_c}(^1P_1^{[8]})\rangle
+d\hat{\sigma}(^1S_0^{[8]})\langle O^{\eta_c}(^1S_0^{[8]})\rangle, \label{eqn:dsigma}
\eea
where we have abbreviated the SDCs $\hat{\sigma}(e^+e^-\rightarrow n+X)$ as $\hat{\sigma}(n)$.
The subscript $lo$ is used to distinguish the QCD LO SDC from the one up to the order of $\alpha^2\alpha_s^2$.

\section{Numerical results and discussions}

Having generated all the needed FORTRAN source using the FDC system~\cite{Wang:2004du},
we start to perform the numerical calculation.
The global choice of the parameters are listed as follows:
The QED and QCD coupling constants are $\alpha=1/137$ and $\alpha_s(3\gev)=0.26$, respectively.
The colliding energy is fixed at $10.6\gev$, which corresponds to the B-factory and Super B-factory experiments.
At this energy, the diagrams involving a $Z$-boson propagator are greatly suppressed.
Therefore we only consider the diagrams in which the electron and positron annihilate into a virtual photon.
We employ the LDMEs obtained in Ref.~\cite{Zhang:2014ybe, Sun:2015pia} as our default choice.
The values of them are also presented below.
\bea
&&\langle O^{\eta_c}(^1S_0^{[1]})\rangle=(0.215\pm0.135)\gev^3, \NO \\
&&\langle O^{\eta_c}(^3S_1^{[8]})\rangle=(0.78\pm0.34)\times 10^{-2}\gev^3, \NO \\
&&\langle O^{\eta_c}(^1S_0^{[8]})\rangle\approx\frac{1}{3}\langle O^{J/\psi}(^3S_1^{[8]})\rangle=0.35\times 10^{-2}\gev^3, \NO \\
&&\langle O^{\eta_c}(^1P_1^{[8]})\rangle\approx3\langle O^{J/\psi}(^3P_0^{[8]})\rangle=5.8\times 10^{-2}\gev^3.
\eea
The last equation has implicated a redefinition of the P-wave LDMEs by the following equation
\be
\langle O^H(^{2S+1}P_J^{[n]})\rangle=\langle O^H(^{2S+1}P_J^{[n]})\rangle_{BBL}/m_c^2, \label{eqn:redef}
\ee
where we use the subscript "BBL" to denote the definition in Ref.~\cite{Bodwin:1994jh}.
Our P-wave SDCs are also redefined by multiplying $m_c^2$ accordingly.

\subsection{Total Cross Sections}

In the following discussions, we use $\sigma(n)$ to abbreviate the contribution of the channel $n$ to the cross section up to the order we keep in our calculation.
For the $^1S_0^{[8]}$ channel, we are also interested in the significance of the QCD corrections,
thus we assign the LO results a distinct name, $\sigma_{lo}(^1S_0^{[8]})$.

Then we can obtain the total cross sections for each channel,
while we choose $m_c=1.5\gev$, $\mu_r=2m_c$ as a default input.
Although the uncertainties of the LDMEs for $^1S_0^{[1]}$ and $^3S_1^{[8]}$ are huge,
the SDCs for the two channels are so small that these contributions are almost negligible,
so, we do not count these uncertainties and just adopt the central value of them.
The results are listed in TABLE\ref{table:tcs}.
One can easily find that the CS contribution, although enhanced by the LDME,
is almost 50 times smaller than the CO one.
This is quite different from the $J/\psi$ case,
in which both the CS and CO contributions are significant.
Accordingly, this process can serve as a good laboratory to test NRQCD.
Another interesting feature of this process is that the $^1S_0^{[8]}$ channel dominates the total cross section,
while the other two CO channels are almost one order of magnitude smaller.
Despite the exploration of numerous processes,
we have not found an example as clean as this one, for the determination of the LDME $\langle O^{\eta_c}(^1S_0^{[8]})\rangle$.

\begin{table}[ht]
\begin{center}
\begin{tabular}  {c c c c c c c c}
 \hline
 \hline
 ~$n$&$^1S_0^{[1]}$~&~$^1S_0^{[8]}$(LO)~&~$^1S_0^{[8]}$~&~$^3S_1^{[8]}$~&~$^1P_1^{[8]}$~&~CO~&~total~\\
 \hline
 ~$\sigma(n)(\mbox{pb})$~&~0.0021~&~0.043~&~0.080~&~0.0128~&~-0.0032~&~0.090~&~0.092~\\
  \hline
\end{tabular}\\
\caption{
The total cross section for the process $e^+e^-\rightarrow\eta_c+$light hadrons.
The results contributed by each channel are also presented.
}\label{table:tcs}
\end{center}
\end{table}

We also need to study the $\mu_r$ and $m_c$ dependence of the total cross section,
which implicates the convergence of the perturbative expansion at a fixed order.
Before we present the numerical results, we need to address the dependence of the LDMEs on the scales.
As Ref.~\cite{Jia:2014jfa} pointed out, the LDMEs do not depend on $\mu_r$,
which is a direct conclusion of the equation
\be
\frac{\partial\langle O^H(n)\rangle}{\partial\mu_r}=0.
\ee
However, as $m_c$ varies its value, the LDMEs scale as~\cite{Bodwin:1994jh, Cho:1995vh}
\be
\langle O^{\eta_c}(n)\rangle\propto m_c^3. \label{eqn:mcdep}
\ee
Note that we have redefined the P-wave LDMEs in Eq.(\ref{eqn:redef}).
Since all the LDMEs we used in this paper are obtained at a fixed value of $m_c$,
we need to take the scaling in Eq.(\ref{eqn:mcdep}) into account in our numerical study.

\begin{figure}[htbp]
\centering
\includegraphics[width=0.8\textwidth]{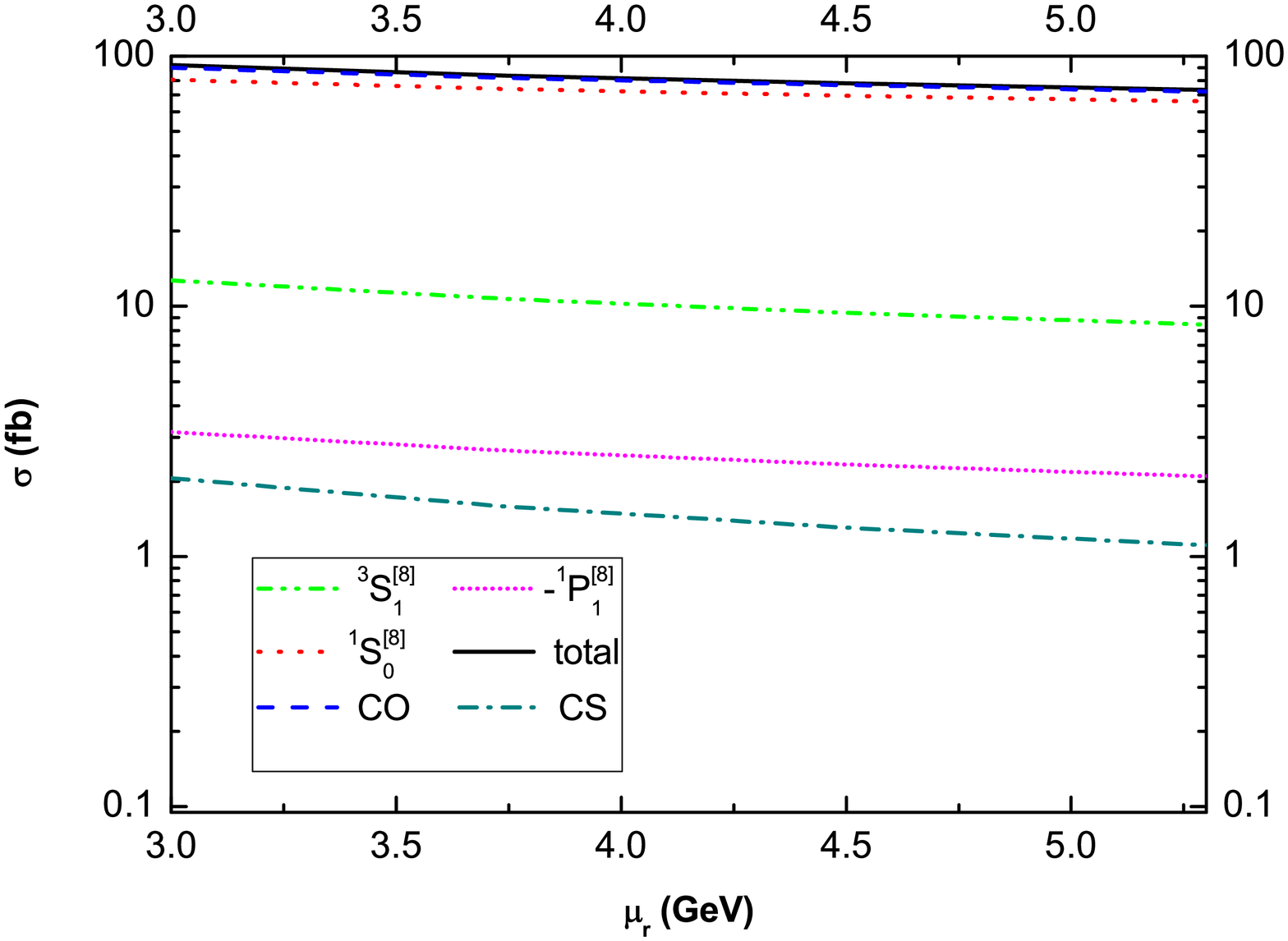}
\caption{
The $\mu_r$ dependence of the total cross sections for the process $e^+e^-\rightarrow\eta_c$+light hadrons.
}\label{fig:murdep}
\end{figure}

The $\mu_r$ dependence of the total cross sections is presented in Fig.\ref{fig:murdep},
where $m_c=1.5\gev$ is fixed.
One can observe that as $\mu_r$ varies from $2m_c=3\gev$ to $\sqrt{s}/2=5.3\gev$,
the total cross section slopes down from $92\mathrm{fb}$ to $70\mathrm{fb}$.
And the CS contribution decreases from $2\mathrm{fb}$ to about $1.3\mathrm{fb}$.
This dependence is comparable with the process $e^+e^-\rightarrow J/\psi+X$~\cite{Ma:2008gq, Gong:2009kp, Gong:2009ng},
which indicates the convergence of the $\eta_c$ production process might not be too bad.

\begin{figure}[htbp]
 \centering
 \includegraphics[width=0.8\textwidth]{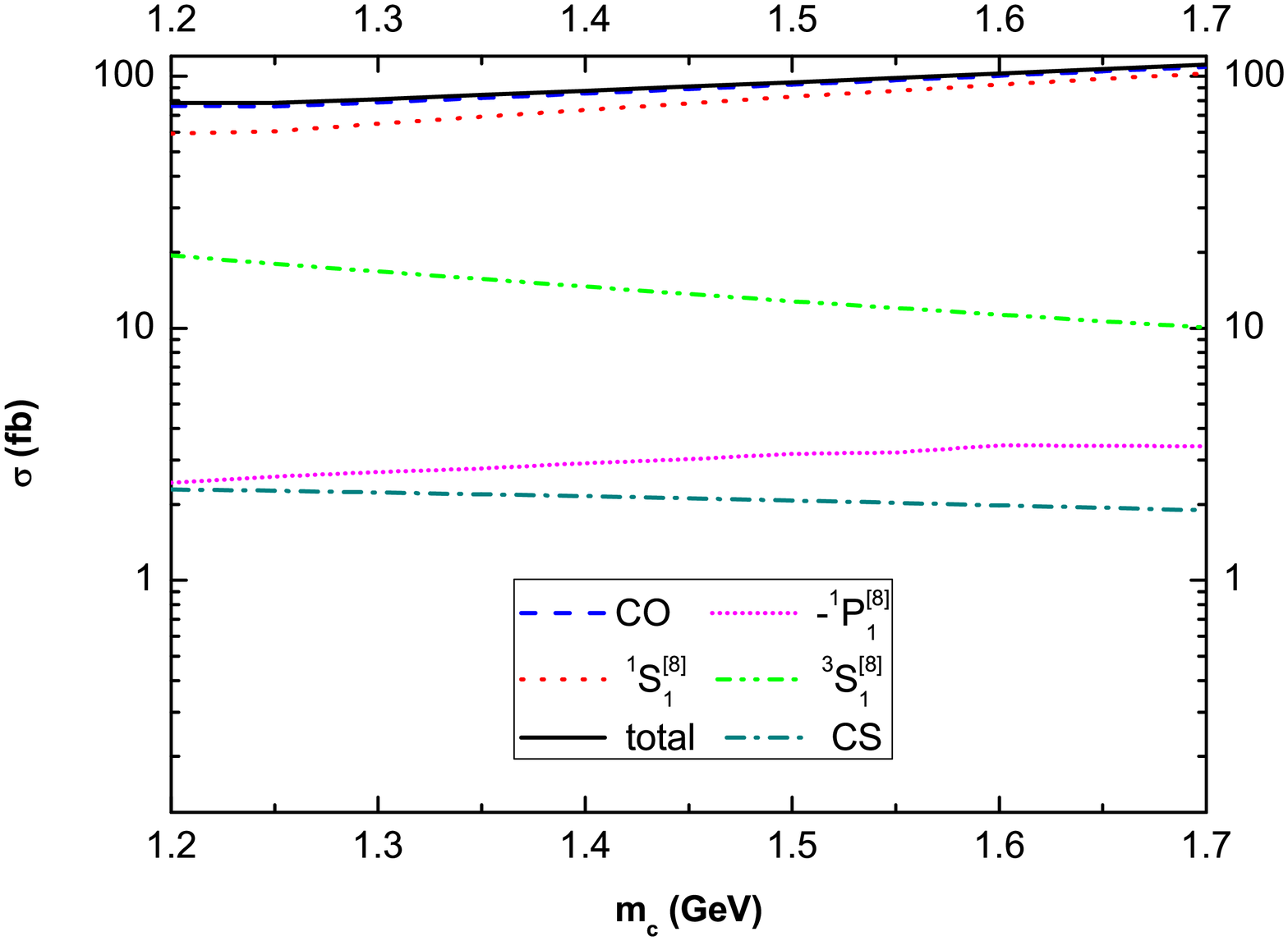}
 \caption{
The $m_c$ dependence of the total cross sections for the process $e^+e^-\rightarrow\eta_c$+light hadrons,
where the scaling in Eq.(\ref{eqn:mcdep}) are taken into account.
}\label{fig:mcdep}
\end{figure}

In Fig.\ref{fig:mcdep}, we present the $m_c$ dependence of the total cross sections,
where $\mu_r=3.0\gev$ is fixed and the scaling in Eq.(\ref{eqn:mcdep}) has been taken into account.
Apparently, $m_c$ dependence for the process we study in this paper is even milder than
that for the $J/\psi$ production processes studied in Ref.~\cite{Ma:2008gq, Gong:2009kp, Gong:2009ng}.

These results suggest it is trustable that the $\eta_c$ production in association with light hadrons at B-factories is dominated by the $^1S_0^{[8]}$ channel.
Accordingly, this experiment can provide an excellent opportunity for the test of the CO mechanism.

\subsection{Angular Distribution}

\begin{figure}[htbp]
 \centering
 \includegraphics[width=0.8\textwidth]{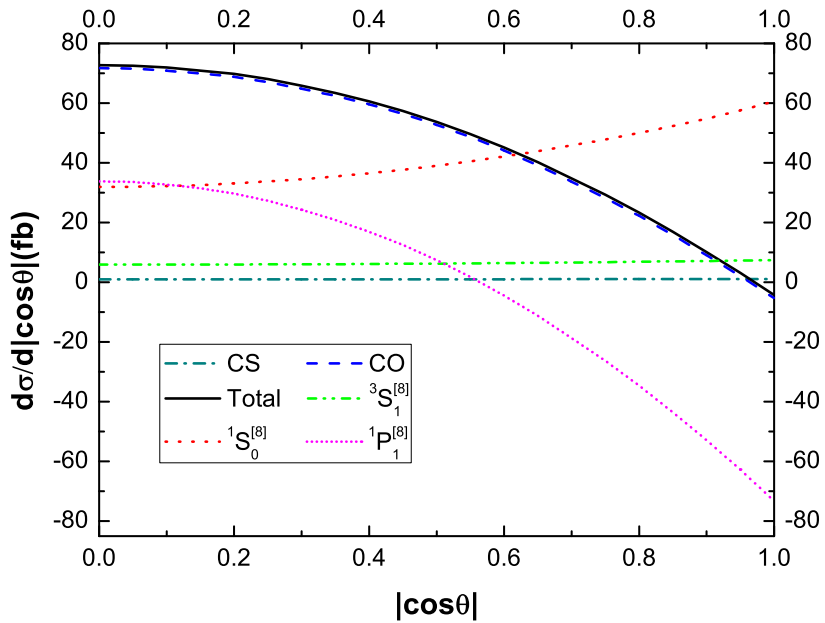}
 \caption{
The differential cross sections for the process $e^+e^-\rightarrow\eta_c$+light hadrons with respect to $\mathrm{cos}\theta$.
}\label{fig:angdist}
\end{figure}

We present the angular distribution of the $\eta_c$ production at B-factories in Fig.\ref{fig:angdist}.
It is also the first time the angular distribution of the $c\bar{c}(^1S_0^{[8]})$ state production in $e^+e^-$ annihilation at QCD NLO is given.
We recall that the angular distribution for the CS contribution to the process $e^+e^-\rightarrow J/\psi+gg$ given in Ref.~\cite{Gong:2009ng} is flat,
while the Belle data~\cite{Pakhlov:2009nj} goes upward as $\mathrm{cos}\theta$ increases.
Interestingly, the $^1S_0^{[8]}$ channel, which also contributes to the $J/\psi$ production, has the same $\mathrm{cos}\theta$ behaviour as the Belle data.
This might indicate the existence of the CO contributions in the $J/\psi$ production process at B-factories.

According to Fig.\ref{fig:angdist}, the $\mathrm{cos}\theta$ distribution is also dominated by the CO channels.
However, the $^1P_1^{[8]}$ contribution is, yet, not negligible;
it completely changes the behaviour of the differential cross section,
even though after integrating out $\mathrm{cos}\theta$ it turns out to be almost zero.
The differential cross section with respect to $\mathrm{cos}\theta$ within the NRQCD framework is downward going.
This kind of behaviour can be regarded as the most distinct signal for the CO mechanism.

One might notice the differential cross section turns out to be negative near the point $\mathrm{cos\theta}=1$.
This is not a severe problem as it seems to be.
First of all, up to QCD NLO, the terms we keep in the perturbative expansion is NOT a perfect square;
the inclusion of the higher-order terms can make the results positive.
Alternatively, one can tune the scales to achieve better results.
Actually, these two operations have the same basis,
since the uncertainty brought in by the different choices of the scales is anyway a higher-order effect.

\subsection{$\mu_\Lambda$ dependence}

To study the convergence of the perturbative expansion, we also need to observe the $\mu_\Lambda$ dependence of the cross sections.
Here we focus on two questions. 1) Does a different choice of $\mu_\Lambda$ change the behavior of the angular distribution?
2) Does the differential cross section near the point $\mathrm{cos}\theta=1$ always lie below 0?

As is indicated by Eq.(\ref{eqn:dsigma}), $\mu_\Lambda$ independence requires
\be
d\hat{\sigma}(^1S_0^{[8]})\propto\alpha_sd\hat{\sigma}_{lo}(^1S_0^{[8]})
\ee
at any value of $\mathrm{cos}\theta$.
In this case, when the value of $\mu_\Lambda$ varies,
one can preserve the differential cross section results by tuning the value of $\langle O^{\eta_c}(^1S_0^{[8]})\rangle$.

Here we define
\be
r=\frac{d\hat{\sigma}(^1S_0^{[8]})}{\alpha_sd\hat{\sigma}_{lo}(^1S_0^{[8]})}, \label{eqn:r}
\ee
which is slightly different from the definition provided in Ref.~\cite{Wang:2014vsa, Jia:2014jfa}.
If $r$ is a constant with respect to $\mathrm{cos}\theta$,
when $\mu_\Lambda$ is changed into $\mu'_\Lambda$,
to make the cross section invariant, the $^1S_0^{[8]}$ LDME should be
\be
\langle O^{\eta_c}(^1S_0^{[8]})\rangle\rightarrow\langle O^{\eta_c}(^1S_0^{[8]})\rangle
+\frac{1}{9\pi m_c^2r}\frac{N_c^2-4}{N_c}\mathrm{ln}(\frac{\mu_\Lambda'^2}{\mu_\Lambda^2})\langle O^{\eta_c}(^1P_1^{[8]})\rangle. \label{eqn:ldmemulam}
\ee
This is also consistent with the renormalisation group equation
\be
\mu_\Lambda\frac{\partial\langle O^{\eta_c}(^1S_0^{[8]})\rangle}{\partial\mu_\Lambda}=\frac{2\alpha_s}{9\pi m_c^2}\frac{N_c^2-4}{N_c}\langle O^{\eta_c}(^1P_1^{[8]})\rangle,
\ee
once the perturbative expansion reaches good convergence at LO.
In this case, $r$ is approximately $1/\alpha_s$.

\begin{figure}[htbp]
 \centering
 \includegraphics[width=0.8\textwidth]{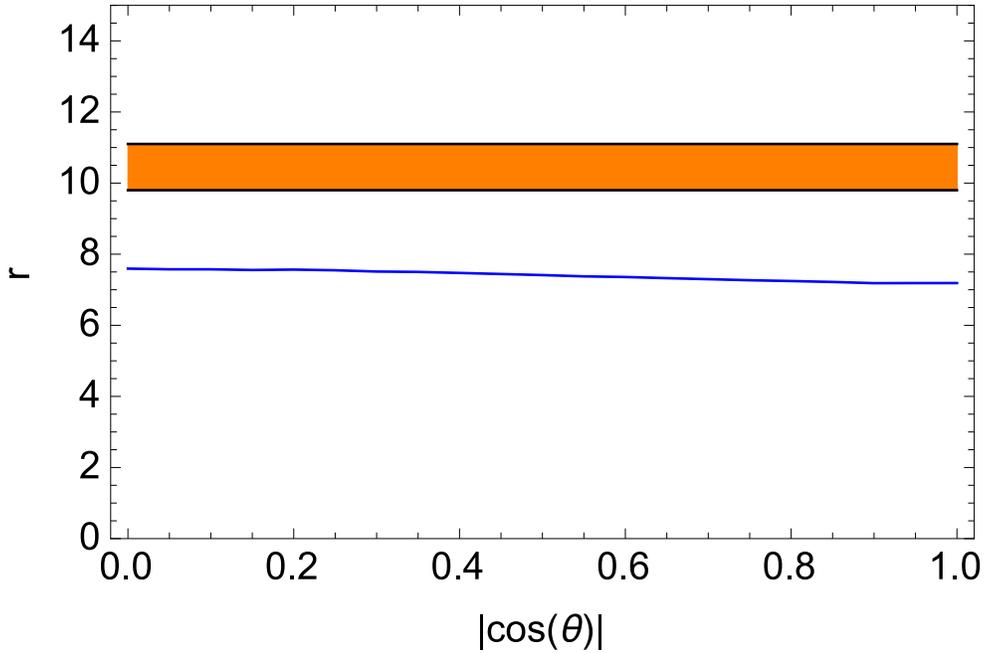}
 \caption{
The value of $r$ defined in Eq.(\ref{eqn:r}) as a function of $\mathrm{cos}\theta$.
The shaded area correspond to range of $r_J$ obtained in Ref.~\cite{Jia:2014jfa}.
}\label{fig:r}
\end{figure}

\begin{figure}[htbp]
 \centering
 \includegraphics[width=0.8\textwidth]{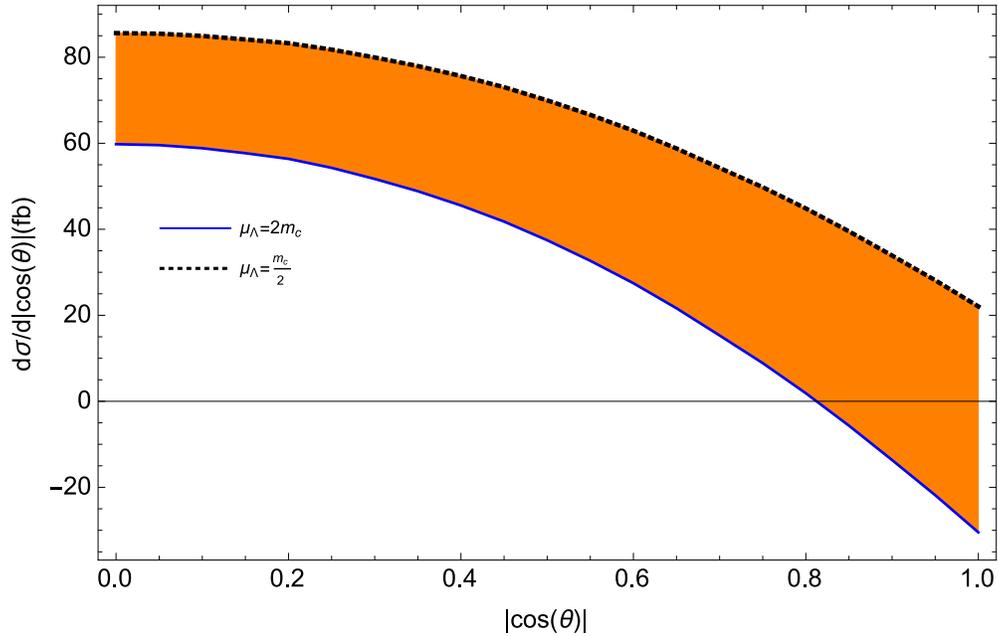}
 \caption{
The angular distribution of $\eta_c$ production in association with light hadrons at B-factories.
The upper and lower bounds of the band correspond to $\mu_\Lambda=m_c/2$ and $\mu_\Lambda=2m_c$, respectively.
}\label{fig:angdistmulam}
\end{figure}

However, the LDMEs are obtained through the fit of the $J/\psi$ data.
If we denote the value of $r$ for the $c\bar{c}(^3S_1^{[8]})$ hadroproduction as $r_J$, namely
\be
r_J=\frac{d\hat{\sigma}(^3S_1^{[8]})}{\alpha_sd\hat{\sigma}_{lo}(^3S_1^{[8]})}, \label{eqn:rj}
\ee
the LDMEs for the $J/\psi$ production also satisfy Eq.(\ref{eqn:ldmemulam}) once replacing $r$ by $r_J$:
\be
\langle O^{J/\psi}(^3S_1^{[8]})\rangle\rightarrow\langle O^{J/\psi}(^3S_1^{[8]})\rangle
+\frac{1}{\pi m_c^2r_J}\frac{N_c^2-4}{N_c}\mathrm{ln}(\frac{\mu_\Lambda'^2}{\mu_\Lambda^2})\langle O^{\eta_c}(^3P_0^{[8]})\rangle, \label{eqn:ldmesc}
\ee
where a factor of 9 is multiplied to compensate the difference between the LDME for $J/\psi$ and $\eta_c$.
Note that $\hat{\sigma}$ in Eq.(\ref{eqn:rj}) represents the SDC for the hadroproduction of the corresponding intermediate state.
The value of $r_J$ ranges from 9.8 to 11.1, as is obtained in Ref.~\cite{Jia:2014jfa}.
In Fig.\ref{fig:r}, we can see that the value of $r$ is quite below that of $r_J$ (the shaded area).
We adopt the central value of $r_J$, namely $r_J=10.5$, and employ Eq.(\ref{eqn:ldmesc}) to obtain the LDMEs at different values of $\mu_\Lambda$.
Even though $r$ is almost a constant with respect to $\mathrm{cos}\theta$,
having $r\neq r_J$, the cross sections for the process we study in this paper still depend on $\mu_\Lambda$.
To illustrate the uncertainties brought in by $\mu_\Lambda$,
we present the band corresponding to the range $\frac{m_c}{2}<\mu_\Lambda<2m_c$ in Fig.\ref{fig:angdistmulam}.
One can find that for $\mu_\Lambda=\frac{m_c}{2}$, the differential cross section is already positive in the whole $\mathrm{cos}\theta$ range.

\section{Summary}

In this paper, we studied the $\eta_c$ associated production with light hadrons in $e^+e^-$ collisions at the B-factory energy.
This process serves as the best device to test the CO mechanism.
We found that the CS contributions are almost negligible,
while the $^1S_0^{[8]}$ channel dominates the total cross section.
The $^1P_1^{[8]}$ channel almost vanishes in the total cross section calculation,
however, proves to be very important for the angular distribution behaviour.
The angular distribution turns out to be downward going when all the CO channels are counted,
which is one of the most distinct signal for the CO mechanism.
We also studied the $\mu_r$, $m_c$ and $\mu_\Lambda$ dependence.
It was found that these dependences are even milder than those for the processes $e^+e^-\rightarrow J/\psi+X$ at the same colliding energy.
We also presented the first study on the angular distribution of $c\bar{c}(^1S_0^{[8]})$ production at the B-factories,
which might be useful for the understanding of the angular distributions of the $J/\psi$ production measured by Belle.

\section{Acknowledgments}

This work is supported by the National Natural Science Foundation of China (Nos.~11405268, Nos.~10925522 and Nos.~11021092).


\end{document}